\documentclass[aps,prl,floatfix,epsfig,twocolumn,showpacs,preprintnumbers]{revtex4}
\usepackage{amsmath,amssymb,graphicx,bm,epsfig}

\renewcommand{\a}{\alpha}
\renewcommand{\b}{\beta}

\newcommand{\vK}{{\mathbf{K}}}
\newcommand{\vq}{{\mathbf{q}}}
\newcommand{\vQ}{{\mathbf{Q}}}

\newcommand{\vS}{{\mathbf{S}}}

\begin{document}

\title{Avoided Quantum Criticality near Optimally Doped High  Temperature Superconductors}

\author{Kristjan Haule and Gabriel Kotliar}

\affiliation{Department of Physics, Rutgers University, Piscataway, NJ 08854, USA}

\begin{abstract}
We study the crossover  from the  underdoped to the overdoped
regime in the t-J model.
The underdoped regime is dominated by the superexchange
interaction, locking the spins into singlets which weakly
perturbe coherent charge carriers. In the overdoped, large carrier
concentration regime, the Kondo effect dominates resulting in
spin-charge composite quasiparticles which are also coherent.
Separating these two Fermi liquid regimes, there is a critical
doping where superexchange and Kondo interaction balance each
other, bringing the system close to a local quantum critical
point near  the point of maximal superconducting transition
temperature. At this point, particle hole symmetry is dynamically
restored and physical quantities such as the optical
conductivity, exhibit power law behaviour at intermediate
frequencies as observed experimentally. Quantum  criticality is
avoided by the onset of superconductivity.
\end{abstract}
\date{\today}
\maketitle

The  high temperature  superconductors   have a normal state
which is  not well described by  the standard theory of metals
\cite{Lee}. Several anomalies are well established
experimentally: a) The normal state close to the doping level
with the highest transition temperature is highly incoherent and
does not fit the standard Fermi liquid picture, b) at small
doping the normal state has a pseudogap, c) the optimally doped
regime exhibits anomalous power laws in transport quantities and
in the optical conductivity in an intermediate frequency range.
This  observation has lead to the  suggestion that the
superconducting dome covers a critical point \cite{Tallon}, d) The
superconducting state not only exhibits coherent Bogolubov
quasiparticles and charge collective modes, but also a very sharp
spin mode or "40 meV resonance". The role of this mode as well as
of the putative quantum critical fluctuations in a fundamental
theory of cuprate supercondcutivity are  important issues.

In this letter we present a coherent account of these observations
based on a Cluster Dynamical Mean Field Theory analysis of the t-J
model. This approach allows us to  access  the regime of
intermediate temperatures in the doping regime between underdoped
to overdoped system.  In addition, the approach is formulated in
terms of pseudoparticles, representing  collective excitations of
the system in an enlarged Hilbert space. This formulation provides
a clear interpretation of the numerical results, providing a
transparent physical interpretation.



We start from the the t-J model which was proposed by P.W. Anderson as
a minimal model to describe the cuprate superconductors
\begin{equation}
  H = -\sum_{ij\sigma} t_{ij}c_{i\sigma}^\dagger c_{j\sigma}+
  \frac{1}{2}\sum_{ij}\mathbf{S}_i\mathbf{S}_j .
\end{equation}
It contains two terms; one describes the kinetic energy which
delocalizes the holes introduced by doping, and a spin spin
interaction.
A constraint of no double occupancy must be enforced.

We used the extended  Dynamical  Cluster Approximation
\cite{Hettler} combined with the slave particle approach
\cite{HaulefU} which have been very useful in the theory of
strongly correlated electrons. The cluster approach maps the full
many body problem, into a a set of local degrees of freedom (in
this case a two by two plaquette) which are treated exactly and a
bath which is treated self consistently.

The coupling of the cluster to the medium, which simulates the
rest of the lattice, causes the cluster eigenstates to decay in
time therefore their spectral functions carry nontrivial
frequency dependence and important information about
various physical processes
such as the RKKY interactions, the  Kondo effect and d-vawe
superconductivity. To study the dynamics of the cluster states,
we diagonalized the cluster, i.e.,
$H_{cluster}|m\rangle=E_m|m\rangle$, and assigned to each cluster
eigenstate a pseudoparticle $|m\rangle=a_m^\dagger|0\rangle$. The
original problem can be exactly expressed in terms of
pseudoparticles $a_m$ with the only non-quadratic term of the
converted Hamiltonian being the hybridization between the cluster
and the medium. This cubic hybridization term is taken into
account by the self-consistent resummation of the non-crossing
diagrams in the spirit of the well known Non-Crossing
approximation to the Anderson impurity model. Since the Hilbert
space of the new base of pseudoparticles accesses some
non-physical states, one needs to project them out with the
constraint expressing the completeness of the cluster eigenbasis.
First we calculate the pseudoparticle self energies and the
pseudoparticle Greens functions. Finally, physical observables
like spectral function or susceptibilities can be calculated by a
weighted sum of the convolution of the pseudoparticle spectral
functions. The following matrix elements
$(F^{\vK\dagger})_{nm}=\langle n|c_\vK^\dagger|m\rangle$ and
$(\vS_Q)_{nm}=\langle n|\vS_Q|m\rangle$ give the bare weight of
the electron spectra and the spin response in terms of
pseudoparticles
\begin{eqnarray}
  G_\vK(i\omega)=-\sum_{nmn'm'}(F^\vK)_{m'n'}C_{m'n'nm}(i\omega)(F^{\vK\dagger})_{nm}\\
  \chi^{\a\b}_\vQ(i\omega) = \sum_{nmn'm'}(S_\vQ^\alpha)_{m'n'}C_{m'n'nm}(i\omega)(S_{-\vQ}^\beta)_{nm}
\end{eqnarray}
with
\begin{eqnarray}
  C_{m'n'nm}(i\omega)=T\sum_{i\epsilon}G_{n'n}(i\epsilon)G_{mm'}(i\epsilon-i\omega)
\end{eqnarray}
Here $G_\vK$, $G_{nn'}$ and $\chi_\vQ$ are electron Green's
function, pseudoparticle Green's function and spin
susceptibility, respectively. This is the central equation of the
approach, relating observables to the pseudoparticle spectral
functions plotted in Fig.~\ref{pseudop}. More details of the
method can be found in \cite{Haule}.
It has been shown that the cluster approach successfully describes
many properties of the high temperature superconductors
\cite{Hettler,Haule,Civelli}.  In this letter we show that it
also unravels the origin of the apparent criticality observed
near critical doping, and its elimination below the
superconducting state.

Physically pseudoparticles
represent coarsed grained versions of the important many body
excitations including fermionic quasiparticles and bosonic collective
modes. They have quantum numbers describing their spin, number of
particles, (which divided by the cluster size, give the density), and
a coarsed grained momentum. In momentum space, the course graining for
the smallest cluster divides the Brillouin zone into four different
regions, namely an inert patch around $(0,0)$ point, the second patch
centered around $(\pi,\pi)$ being highly non-Fermi liquid like at
small doping and finally $(\pi,0)$ and $(0,\pi)$ regions which contain
the Fermi surface of the model in the underdoped and slightly
overdoped regime and are therefore most relevant for thermodynamical
and transport properties.

\begin{figure}
\includegraphics[width=0.8\linewidth]{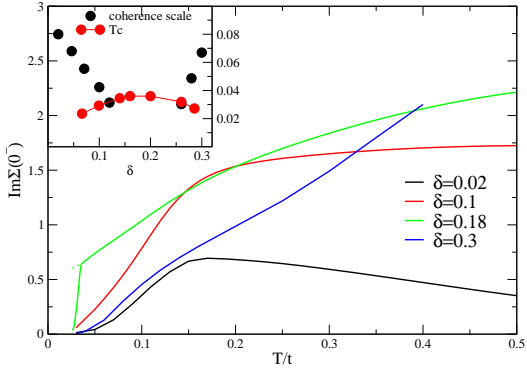}
\caption{
  The cluster $(\pi,0)$ self-energy at zero frequency as a
function of temperature for few doping levels. The inset shows the
estimation of the coherent scale in the normal state of the t-J model
(black dots) and transition temperature to superconducting state (red
dots).
}
\label{phased}
\end{figure}

The first indication for an underlying criticality near optimal
doping comes from the evaluation of the electron scattering rate
obtained from the imaginary part of the electron self-energy as a
function of temperature for few different doping levels.  As
described in Fig.~\ref{phased} both at large and small doping the
scattering rate is small as expected for a Fermi liquid.
Remarkably it becomes very large in the region near optimal
doping when the critical temperature is maximal. The transition
to the superconducting state severely reduces the scattering rate
eliminating the traces of the underlying critical behaviour,
hence the name avoided quantum criticality. A coherence scale,
estimated from the scattering rate, is plotted in the inset of
Fig.~\ref{phased} and shows it tends to  vanish close to  the
point of maximal superconducting transition temperature.

\begin{figure}
\includegraphics[width=0.8\linewidth]{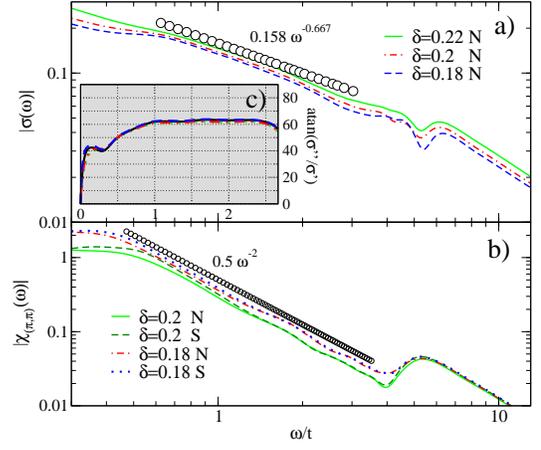}
\caption{ a) The absolute value of complex optical conductivity
$\sigma(\omega)$ is proportional to $\omega^{-2/3}$ in the
intermediate frequency region for optimally doped system. b) The
absolute value of the $(\pi,\pi)$ spin sussceptibility is proportional
to $\omega^{-2}$. c) The angle, calculated as
$\arctan(\mathrm{Im}(\sigma)/\mathrm{Re}(\sigma)$) is approximately
$\pi/3$ since $\sigma(\omega)\sim(-i\omega)^{-2/3}$.  In the legend, N
stand for normal state and S for superconducting state.}
\label{powerlaws}
\end{figure}

Additional  evidence for the quantum criticality is the emergence
of power-law behaviours of the response functions.
As shown in Fig.~\ref{powerlaws},  the optical conductivity has
an approximate power law with an  exponent  $2/3$. The same
powerlaw is realized in the one-electron self-energy  while the
spin susceptibility is proportional to $\chi_{\pi,\pi}\propto
\omega^{-2}$ in the same frequency range. Experimentally, it was
found that the optical conductivity is proportional to the
$(-i\omega)^{-0.65}$ in the intermediate frequency regime
\cite{VanderMarel}.

\begin{figure}
\includegraphics[width=0.8\linewidth]{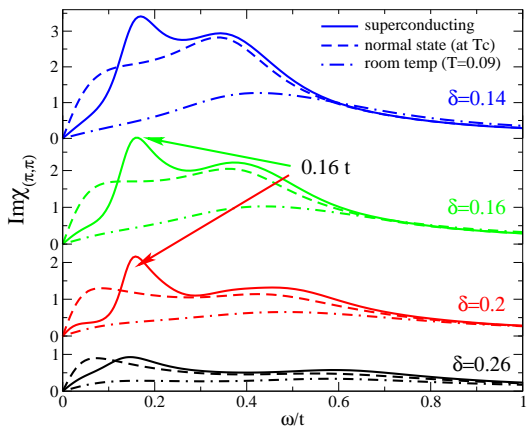}
\caption{
  The dynamical spin susceptibility at $\vq=(\pi,\pi)$ for few
  different doping levels and three different temperatures:
  superconducting state, normal state at the transition temperature
  and at room temperature. The pronounced peak is formed in SC
  state at $0.16t\approx 48\,\mathrm{meV}$ and a broad peak in normal state is
  around $100-140\,\mathrm{meV}$.
  Susceptibility at normal temperture is
  much smaller and the peak moves to higher frequencies.
  The resonance is strongest at the  optimally
  doped system.  It disappears quickly in the overdoped site and
  somewhat more slowly  in the underdoped side.
}
\label{susc40}
\end{figure}

Quantum criticality is avoided when the electrons condense forming
d-wave pairs. The electron scattering rate is dramatically
reduced (see Fig.~\ref{phased}) and a V-shaped gap opens in the
local one-electron density of states.
The particle-hole response at $(\pi,\pi)$
is severely reduced for frequencies  below the superconducting
gap and a sharp
resonance appears in the gap. A sharp resonance peak in the spin
susceptibility was
observed  in bilayer cuprates YBCO and Bi2212 and
in
singe layer Tl2201.
It  scales with doping like $5 T_c$ and does not depend on
temperature.

Our method gives a pronounced peak in the $(\pi,\pi)$ spin response at
frequency $0.16t$ in the optimally doped regime, as shown in
Fig.~\ref{susc40}. The peak position is temperature independent and
taking into account the maximal transition temperature in our model
$T_{c-max}\sim 0.035t$ the resonance energy is $4.6 T_c$ in good
agreement with experiments.

In addition we see a broader peak around $0.35-0.45t$ extending to
very high frequencies of order of $t\approx 300\,\mathrm{meV}$
which also gains some weight upon condensation and represents an
important   contribution  to the exchange energy difference
between superconducting and normal state. This exchange energy
gives far the largest contribution to the condensation energy
\cite{Haule}.

Our method averages the momentum dependence over $1/4$ of the
Brillouin zone centered at $(\pi,\pi)$ therefore it is reasonable
to compare our results with the $\vq$ integrated susceptibility
from Ref.~\onlinecite{Dai} where both the $35\,\mathrm{meV}$
resonant peak as well as broader peak around $75\,\mathrm{meV}$
extending up to $220\,\mathrm{meV}$ was observed.

\begin{figure}
\includegraphics[width=0.9\linewidth]{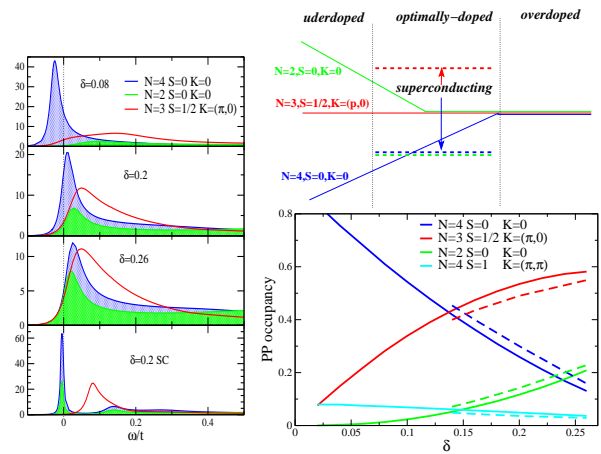}
\caption{ left: Pseudoparticle spectral functions for the
three most important pseudoparticles: ground states for N=4, N=3 and
N=2 sectors. Right-top: Sketch of pseudoparticle treshold energies
which can be interpreted as the effective many-body levels in normal
and superconducting state. Right-bottom: Pseudoparticle occupancies
versus doping for most important pseudoparticles. The full lines
correspond to the normal state while the dashed lines correspond to
the superconducting state.  }
\label{pseudop}
\end{figure}

{\it Pseudoparticle interpretation}: These results can be
understood in terms of the evolution of the pseudoparticle
spectral functions (spectral functions of the eigenstates of the
plaquette) depicted in Fig.~\ref{pseudop}. Remarkably, out of the
large number of pseudoparticles which are involved in our
calculation ($3^4$) only four distinct pseudoparticles are very
important, having the largest weight in the ground state and
containing more than 95\% of the spectral weight of the low energy
correlation functions. Fig.~\ref{pseudop}a shows the evolution of
the three most important pseudoparticle spectral functions from
the underdoped to the  overdoped regime.

{\it Underdoped Regime:} At small doping, the singlet state with
one particle per site and zero momentum (N=4,S=0,K=0) (half filled
singlet) dominates since it has far the lowest treshold energy, is
very strongly peaked at the treshold energy, and has the largest
occupancy as shown in Fig.~\ref{pseudop}c,. It describes a system
locked in a short range singlet state as a consequence of the
strong superexchange interaction.  The electron spectral function
descrbies the process of addition and removal of an electron from
the system at frequency $\omega$. It is dominated by convolution
of two pseudoparticles with different cluster occupation N and
N+1, or N-1 and with integral in convolution taken between zero
and $\omega$. The necessary condition for a peak of the
one-particle spectral function at the Fermi level is that at
least two pseudoparticle spectral functions share common
threshold and are strongly peaked at the same threshold.  Since
the thresholds of the other pseudoparticles are significantly
shifted with reference to the half filled singlet, a pseudogap
results in the one particle spectra in the underdoped regime.
This gap in threshold energies severely limits the possible decay
processes of the electron
resulting
in a low electronic scattering rate.

{\it The local picture of the overdoped regime:} On the overdoped
side, all three important pseudoparticles (half-filled singlet,
doublet with one hole per plaquette and singlet with two holes
per plaquette) have a power law divergence at the same threshold
frequency at zero temperature (Fig.~\ref{pseudop}a and
Fig.~\ref{pseudop}b) which is a standard signature of the Kondo
effect. Hence, the one particle spectral function develops the
Kondo-Suhl resonance at the Fermi level since the convolution
between the doublet and half-filled singlet (or N=2 singlet)
state is large at low frequency.  The one-particle spectral
function is peaked slightly above the Fermi level. The underdoped
regime is distinctively different since only the half-filled
singlet state dominates while the doublet has very little
spectral weight in the region of the singlet peak.

{\it The Transition Region, Normal State:} In the optimal doped
regime, the Kondo effect and the superexchange  compete giving
rise to a  regime with very large scattering rate and
consequently a small coherence scale. Criticality is the result
of the merging of the
convergence of the
thresholds  of the
pseudoparticles (See Fig 4)  when going from  the underdoped to
the overdoped regime.
Local criticality has been found in models of heavy fermion
systems \cite{Si} and  frustrated magnets \cite{georges}.

Surprisingly this evolution is accompanied by a restoration of
particle hole symmetry in the optimally doped regime.  As shown in
Fig.~\ref{pseudop}b the threshold of the N=2 cluster ground state
and N=3 cluster ground state (doublet) merge first resulting in a
Kondo-like contribution to the electron spectral function. This
contribution is peaked above the Fermi level as usual for the
simple Fermi-liquid state in less than half-filled one band
model. The half-filled singlet however remains the lowest state
in energy and still gives a significant contribution to the
electron spectral function. The later contribution is peaked
below the Fermi level and keeps a pseudogap-like shape. The
result of the two contributions to the electron spectral function
is a restoration of the particle-hole symmetry in the density of
states both in normal and superconducting state at optimal
doping. The particle-hole symmetry of the density of states is
yet another hint of the proximity to the quantum critical point
since it is well known that only this symmetry allows non trivial
criticality in the  two impurity Kondo model \cite{varma}.

{\it Transition into the superconducting state:} The degeneracy
responsible for the strongly incoherent metal with large
scattering rate at the Fermi level is lifted by the
superconductivity avoiding the quantum critical point.
Fig.~\ref{pseudop}a shows that both important singlet
pseudoparticles (for N=4 and N=2) develop a very sharp peak at
the same treshold frequency and, at the same time, their
occupancy increases (see Fig.~\ref{pseudop}c) upon condensation,
indicating that electrons are locked into singlets with zero
momentum. A gap opens between the singlets and doublets which
gives the gap in the one-particle density of states. Because of
this gap in the pseudoparticle thresholds, the large imaginary
part of the electron self-energy does not persist in the
superconducting state (see also Fig.~\ref{phased}). Since the
density of states is composed of two almost equally important
contributions, i.e., the convolution of the doublet with both
singlets (N=4 and N=2), the superconducting gap is almost
particle hole symmetric in the optimally doped regime with
half-width of the order of 0.12t. When the doping value is
changed from its critical value, the asymmetry in the
superconducting density of states appears.  The magnitude of the
assymetry is the same as the asymmetry of the corresponding
normal state spectra and comes from the fact that the occupancy
and therefore importance of the N=4 half-filled singlet exceeds
the importance of the N=2 singlet (see Fig.~\ref{pseudop}c). The
spin susceptibility comes almost entirely from the convolution of
the half-filled singlet with the half-filled triplet
($N=4,S=1,K=(\pi,\pi)$). The later develops a  peak at an energy
$0.16t$ upon condensation which causes the resonance in the spin
susceptibility.

\textit{In conclusion} we showed that at optimal doping, in the t-J
model, there is a collapse of coherence scale, accompanied by a a
large scattering rate, restoration of particle-hole symmetry, which
result in approximate power laws in physical quantities, such as the
optical conductivity ($\sigma\propto\omega^{-2/3}$) and spin
susceptibility ($\chi_{\pi\pi}\propto \omega^{-2}$). Quantum
criticality is avoided when electrons condense into a supercondacting
state which results in a dramatic reduction of scattering rate, the
formation of an energy gap which in turns leads to a resonance in the
spin susceptibility.  A transparent physical interpretation of the
numerical results was given in terms of pseudoparticles which
represent the important many body excitations, fermionic
quasiparticles and bosonic collective modes. This real space picture
complements the momentum space picture which emerges in other studies,
where substantial deformations of the Fermi surface which result in
the formation of Fermi arcs at comparable values of the doping
\cite{Haule,tdstan}. Finding analytic connection between single
particle properties in $k$ space and the collective excitations
presented in this study is an interesting direction for future
research.

\end{document}